\documentclass[pre,twocolumn,preprintnumbers,amsmath,amssymb,showpacs,nofootinbib,floatfix]{revtex4}

\usepackage{graphicx,bm}

\makeatletter
\def\graphicscale{\twocolumn@sw{0.3}{0.4}}
\def\graphicthreescale{\twocolumn@sw{0.3}{0.4}}

\begin{document}

\title{Shape dependence and anisotropic finite-size scaling of the phase coherence \\

of three-dimensional Bose-Einstein condensed gases}

\author{Giacomo Ceccarelli, Francesco Delfino, Michele Mesiti, and
  Ettore Vicari}

\address{Dipartimento di Fisica dell'Universit\`a di Pisa and
  INFN, Largo Pontecorvo 3, I-56127 Pisa, Italy} 

\date{\today}

\begin{abstract}

We investigate the equilibrium phase-coherence properties of
Bose-condensed particle systems, focussing on their shape dependence
and finite-size scaling (FSS).  We consider three-dimensional (3D)
homogeneous systems confined to anisotropic $L\times L \times L_a$
boxes, below the BEC transition temperature $T_c$.  We show that the
phase correlations develop peculiar anisotropic FSS for any $T<T_c$,
in the large-$L$ limit keeping the ratio $\lambda\equiv L_a/L^2$
fixed.  This phenomenon is effectively described by the 3D spin-wave
(SW) theory.  Its universality is confirmed by quantum Monte Carlo
simulations of the 3D Bose-Hubbard model in the BEC phase.  The
phase-coherence properties of very elongated BEC systems, $\lambda\gg
1$, are characterized by a coherence length $\xi_a\sim A_t \rho_s/T$
where $A_t$ is the transverse area and $\rho_s$ is the superfluid
density.

\end{abstract}


\maketitle



\section{Introduction}
\label{intro}

The low-temperature behavior of three-dimensional (3D) bosonic gases
is characterized by the formation of a Bose-Einstein condensate (BEC),
below a finite-temperature BEC phase transition.  The phase coherence
properties of cold atomic gases within the low-temperature BEC phase
have been investigated by several experiments with cold atoms in
harmonic traps, see, e.g.,
Refs.~\cite{Andrews-etal-97,Stenger-etal-99,Hagley-etal-99,BHE-00,
  PSW-01,Dettmer-etal-01,Hellweg-etal-02,Cacciapuoti-etal-03,
  Hellweg-etal-03,Ritter-etal-07}.  The coherence length turns out to
be equal to the condensate size in generic 3D traps.  However, very
elongated harmonic traps may give rise to a substantial phase
decoherence along the longer axial
direction~\cite{PSW-01,Dettmer-etal-01,Hellweg-etal-02,Mathey-etal-10,GCP-12}.

In this paper we study the equilibrium phase-coherence properties of
homogeneous Bose-condensed particle systems, focussing on their shape
dependence and finite-size scaling (FSS).  We show that homogeneous
BEC systems constrained within $L\times L \times L_a$ boxes develop a
peculiar anisotropic FSS (AFSS), in the large-$L$ limit keeping the
ratio $\lambda\equiv L_a/L^2$ fixed.  This AFSS is effectively
described by a 3D spin-wave (SW) theory, providing a quantitative
description of the crossover from 3D box geometries to elongated
effectively 1D systems.  The phase-correlation properties of the 3D SW
theory are expected to be universal, i.e., they are expected to apply
to any condensed particle system, at any temperature below the BEC
transition.  In particular, for very elongated BEC systems
($\lambda\gg 1$) our analysis confirms that the one-particle
correlation function decays exponentially along the axial direction,
with a coherence length proportional to the transverse area.

\begin{figure}
\includegraphics{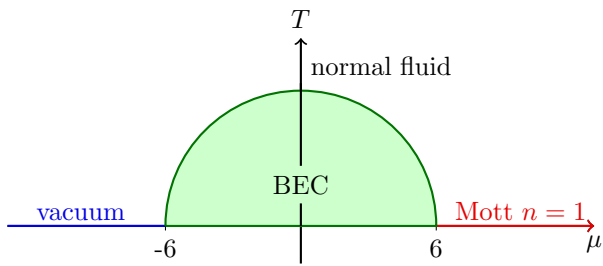}
\caption{(Color online) Sketch of the $T$-$\mu$ (in unit of the
  hopping parameter $t$) phase diagram of the 3D BH model in the
  hard-core $U\to\infty$ limit.  The BEC phase is restricted to a
  finite region between $\mu=-6$ and $\mu=6$.  It is bounded by a BEC
  transition line $T_c(\mu)$, which satisfies $T_c(\mu)=T_c(-\mu)$ due
  to a particle-hole symmetry.  Its maximum occurs at $\mu=0$,
  where~\cite{CN-14} $T_c(\mu=0)= 2.0160(1)$.  At $T=0$ two further
  quantum phases exist: the vacuum phase ($\mu<-6$) and the
  incompressible $n=1$ Mott phase ($\mu>6$).  }
\label{3dphasedia}
\end{figure}

An interesting many-body system showing BEC is the 3D Bose-Hubbard
(BH) Hamiltonian~\cite{FWGF-89}, which models gases of bosonic atoms
in optical lattices~\cite{JBCGZ-98}.  It reads
\begin{eqnarray}
H &=& - t \sum_{\langle ij\rangle} (b_i^\dagger b_j+
b_j^\dagger b_i) + \label{bhm}\\
&+&{U\over 2} \sum_i n_i(n_i-1) - \mu \sum_i n_i\,,
\nonumber
\end{eqnarray}
where $b_i$ is a bosonic operator, $n_i\equiv b_i^\dagger b_i$ is the
particle density operator, and the sums run over the bonds ${\langle
  ij \rangle }$ and the sites $i$ of a cubic lattice.  We set the
hopping parameter $t=1$, so that all energies are expressed in units
of $t$.  The phase diagram of 3D BH models and their critical
behaviors have been much investigated, see
e.g. Refs.~\cite{FWGF-89,CPS-07,CR-12,CTV-13,CN-14,CNPV-15}.  Their
$T$-$\mu$ phase diagram, see for example Fig.~\ref{3dphasedia},
presents a finite-temperature BEC transition line, characterized by
the accumulation of a macroscopic number of atoms in a single quantum
state.  The condensate wave function provides the complex order
parameter of the BEC transition, whose critical behavior belongs to
U(1)-symmetric XY universality class~\cite{PV-02}.  The BEC phase
extends below the BEC transition line.  The phase coherence properties
are inferred from the one-particle correlation function
\begin{eqnarray}
G({\bf x},{\bf y}) \equiv \langle b_{\bf x}^\dagger b_{\bf y}
\rangle.
\label{gbdef}
\end{eqnarray}
Its behavior, and  the related momentum distribution, can
be experimentally investigated by looking at the interference patterns
of absorption images after a time-of-flight period in the large-time
ballistic regime~\cite{BDZ-08}.

We present a numerical analysis of the space-coherence properties of
the 3D hard-core BH model within its BEC phase, by quantum Monte Carlo
(QMC) simulations. The results agree with the AFSS behaviors obtained
within the 3D SW theory, supporting its universality.

The paper is organized as follows.  In Sec.~\ref{3dsw} we present the
3D SW theory that provides an effective description of the long-range
phase correlations within the BEC phase, and show the emergence of a
nontrivial AFSS behavior.  In Sec.~\ref{becphase} we investigate the
space coherence of the 3D hard-core BH model within its BEC phase by
QMC simulations, confirming the universality of the AFSS behaviors
obtained within the 3D SW theory.  Finally, in Sec.~\ref{conclu} we
draw our conclusions.  We also add an appendix containing some details
of our numerical calculations.

\section{3D spin-wave theory}
\label{3dsw}

Within the BEC phase, and for $T\ll T_c$ when the density
$\rho_0=\langle n_0 \rangle$ of the condensate is much larger than the
density of the noncondensed atoms, the particle-field operator of
homogeneous systems can be effectively approximated by~\cite{PSW-01}
$b({\bf x}) = \sqrt{n_0} e^{i\theta({\bf x})}$. Then, the
long-distance modes of the phase correlations are expected to be
described by an effective 3D SW theory for a real phase field
$\theta({\bf x})$, which is invariant under a global shift
$\theta({\bf x})\to \theta({\bf x}) + \varphi$.  The simplest SW
action reads
\begin{equation}
S_{\rm sw} = \int d^dx\; {\alpha \over 2} (\partial_\mu\theta)^2
\label{swa}
\end{equation}
where $\alpha$ plays the role of the superfluid density (as normally
defined phenomenologically,~\cite{FBJ-73} $\alpha\propto \rho_s/T$).
In the framework of the effective field theories~\cite{Weinberg-book},
Eq.~(\ref{swa}) represents the first non-trivial term of a derivative
expansion of the fundamental field $\theta({\bf x})$.  Contributions
of higher-order derivatives are expected to be suppressed in the
long-range correlations. We return to this point later.

Actually, as argued in Ref.~\cite{PSW-01}, the region where the SW
theory effectively describes the long-distance phase correlations is
expected to extend to the whole BEC phase, i.e. for $T\lesssim T_c$,
excluding only the relatively small critical region close to
$T_c$. This is essentially related to the fact that the fluctuations
of the condensate density are suppressed below the critical region
around $T_c$.  Therefore, the two-point function
\begin{equation}
G_{\rm sw}({\bf x-y}) = \langle e^{-i\theta({\bf x})} \,
e^{i\theta({\bf y})} \rangle
\label{twopsw}
\end{equation}
is expected to describe the long-range phase-coherence properties of
particle systems in the whole BEC phase.

We consider the 3D SW model in a finite box of generic shape
$L_1\times L_2 \times L_3$ and periodic boundary conditions (PBCs).
We regularize the theory on a corresponding $L_1\times L_2 \times L_3$
lattice, and write the partition function as
\begin{eqnarray}
 Z &=&  \sum_{\{ n_\mu\}} \int \mbox{D} [\theta] \;
 e^{-\frac{\alpha}{2} \sum_{{\bf x},\mu} 
 (\theta_{\bf x} - \theta_{{\bf x}+\hat \mu} - 2 \pi n_\mu  \delta_{x_\mu L_\mu})^2}
 \nonumber \\
  &= &\sum_{\{ n_\mu\}} W(n_1,n_2,n_3) \int \mbox{D} [\theta]\;
 e^{-\frac{\alpha}{2} \sum_{{\bf x},\mu} (\theta_{\bf x} - \theta_{{\bf x}+\hat \mu})^2}
 \nonumber\\
  &=& \sum_{\{n_\mu\}}W(n_1,n_2,n_3)\; Z_0 ,
 \label{zsw}
\end{eqnarray}
where $\mu=1,\,2,\,3$, $n_\mu\in \mathbb{Z}$, the shift $2 \pi n_\mu
\delta_{x_\mu L_\mu}$ at the boundary takes into account winding
configurations on lattices with PBC, the weights $W(n_\mu)$ are given
by
\begin{equation}
\ln W = - 2 \pi^2 \alpha \left( \frac{L_2L_3}{L_1} n_1^2 +
\frac{L_3L_1}{L_2} n_2^2 + \frac{L_1L_2}{L_3} n_3^2 \right),
\end{equation}
and $Z_0$ is the plain partition function without shift at the
boundaries.  Analogous formulations of the SW theory have been
considered to describe the quasi-long-range order of two-dimensional
U(1)-symmetric systems~\cite{Hasenbusch-05,HPV-05,PS-00,MSS-04}.

The above formulas allow us to compute the helicity modulus along the
three spatial directions, from the response of the system to a phase
twisting $\phi$ along one of the lattice directions~\cite{FBJ-73}. We
obtain
\begin{eqnarray}
&&Y_1 \equiv - \frac{L_1}{L_2 L_3} \left.  \frac{\partial^2 \log
    Z(\phi)}{\partial\phi^2} \right|_{\phi=0} = \label{y1sw}\\ &&=
  \alpha - 4 \pi^2 \alpha^2 \frac{L_2 L_3}{L_1}
  \frac{\sum_{n=-\infty}^\infty n^2 e^{-2 \pi^2 n^2 \alpha L_2L_3/L_1}
  } {\sum_{n=-\infty}^\infty e^{-2 \pi^2 n^2 \alpha L_2L_3/L_1} }
  \nonumber
\end{eqnarray}
and analogously for $Y_2$ and $Y_3$.

The two-point function (\ref{twopsw}) can be written as
\begin{eqnarray}
&&G_{\rm sw}({\bf x-y}) =\langle 
e^{-i (\theta_{\bf x}-\theta_{\bf y})}\rangle_0
\label{correlation1}\\
&&\;\;  \times \frac{\sum_{\{n_\mu\}} W(n_\mu)
\cos\left[2\pi \sum_{\mu=1}^3 {n_\mu (x_\mu-y_\mu)/L_\mu} \right] }
{ \sum_{\{n_\mu\}} W(n_\mu) } ,
\nonumber
\end{eqnarray}
where $\langle . \rangle_0$ denotes the expectation value in a
Gaussian system without boundary shift (with PBC).  Then we use the
relation
\begin{eqnarray} 
&&\langle e^{-i (\theta_{\bf x}-\theta_{\bf y})} \rangle_0 =
\exp[G_0({\bf x},{\bf y})]
\label{correlation2}
\end{eqnarray}
where 
\begin{eqnarray}
G_0({\bf x},{\bf y}) &\equiv& 
\langle (\theta_{\bf x}-\theta_{\bf y})^2 \rangle_0
\label{gauint}\\
&=& \frac{1}{\alpha L_1L_2L_3}
\sum_{{\bf p}\ne {\bf 0}} \frac{\cos[{\bf p}\cdot ({\bf x-y})] - 1} 
{2 \sum_\mu (1 - \cos{p_\mu})}
\nonumber
\end{eqnarray}
with $p_\mu = 2\pi \{ 0,\dots,L_\mu-1 \} / L_\mu$.
Eqs.~(\ref{correlation1}-\ref{gauint}) allow us to compute the $G_{\rm
  sw}({\bf x})$ for any lattice shape and size. The continuum limit is
formally equivalent to the FSS limit, i.e. $L_\mu\to\infty$ keeping
appropriate ratios of the sizes fixed.

As already noted in Ref.~\cite{PS-00}, when we consider large-volume
limits keeping the ratios of the sizes $L_\mu$ finite,
Eq.~(\ref{y1sw}), and the analogous equations for the other
directions, give equal helicity modulus. We obtain $Y_\mu = \alpha$
for any $\mu=1,2,3$. However, nontrivial results are obtained when
considering elongated geometries, in particular when one size scales
as the product of the sizes of the other two directions, giving rise
to an AFSS.

Let us now consider the particular case of anisotropic geometries: we
fix $L_1=L_2=L$ and $L_3=L_a$. More precisely, we consider the AFSS
limit obtained by the large-$L$ limit at fixed ratio
\begin{equation}
\lambda\equiv L_a/L^2 .
\label{defla}
\end{equation}

\begin{figure}[tbp]
\includegraphics*[scale=\graphicscale]{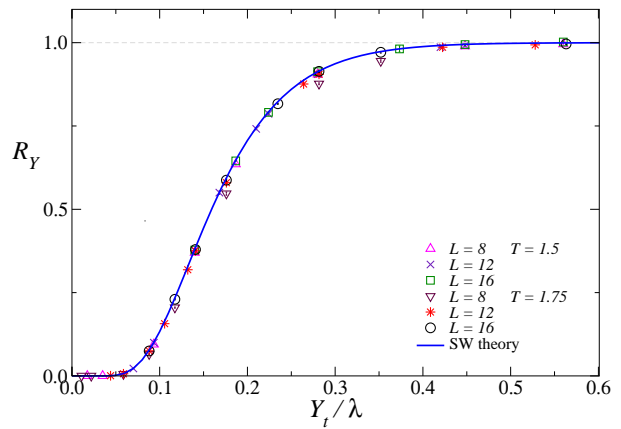}
\caption{(Color online) The ratio $R_Y\equiv Y_a/Y_t$ for anisotropic
  $L^2\times L_a$ lattices with PBC, versus $Y_t/\lambda$ where
  $\lambda\equiv L_a/L^2$. We show the curve (\ref{fyup}) obtained by
  the 3D SW theory (full line), and QMC data for the 3D hard-core BH
  model, for $\mu=0$ and two temperature values, i.e. $T=1.5$ and
  $T=1.75$ (we use the QMC estimates $Y_t\approx 0.280$ and
  $Y_t\approx 0.176$ respectively).  The MC data clearly approach the
  SW AFSS curve with increasing $L$ (the differences get rapidly
  suppressed, apparently as $L^{-3}$).  }
\label{yafig}
\end{figure}

From Eq.~(\ref{y1sw}) we obtain
\begin{eqnarray}
&&Y_t = Y_1 = Y_2 = \alpha ,\label{yt}\\
&&Y_a = Y_3 = \alpha - {4 \pi^2 \alpha^2\over \lambda}
               \frac{\sum_{n=-\infty}^\infty n^2
              e^{-2 \pi^2 n^2 \alpha/\lambda} 
                  }
                   {\sum_{n=-\infty}^\infty   e^{-2 \pi^2 n^2 \alpha/\lambda} 
                                      }.\quad
\label{ya}
\end{eqnarray}
Therefore, the transverse helicity modulus $\Upsilon_t\equiv T Y_t$
can be again identified with the superfluid density. On the other
hand, $Y_a$ varies significantly with increasing $\lambda$, from
$Y_a=\alpha$ for $\lambda\to 0$ to $Y_a\to 0$ for $\lambda \to
\infty$.  Its AFSS can be written as
\begin{eqnarray}
R_Y \equiv Y_a /Y_t = f_\Upsilon(\zeta),\qquad \zeta \equiv Y_t/\lambda,
\label{fssya}
\end{eqnarray}
where 
\begin{eqnarray}
f_\Upsilon(\zeta) 
= 1 + 2  \zeta \, \partial_\zeta \ln \vartheta_3(0|i2\pi\zeta),
\label{fyup}
\end{eqnarray}
and $\vartheta_3(z|\tau)$ is the third elliptic theta
function~\cite{WW-book}.  In particular, $f_\Upsilon(0) = 0$ and
$f_\Upsilon(\infty) = 1$.  A plot of $f_\Upsilon(\zeta)$ is shown in
Fig.~\ref{yafig}.

In the infinite-axial-size $L_a\to\infty$ limit, the phase correlation
$G_{\rm sw}({\bf x})$ along the axial direction is essentially
determined by the first factor of the r.h.s. of
Eq.~(\ref{correlation1}), since axial boundary terms become
irrelevant, as indicated by the vanishing axial helicity modulus.  In
this limit $G_{\rm sw}({\bf x})$ turns out to decay exponentially
along the axial direction:
\begin{equation}
G_{\rm sw}(0,0,z\gg 1) \sim e^{-z/\xi_a}
\label{gooz}
\end{equation}
(apart from a power-law prefactor), where 
\begin{equation}
\xi_a = 2 \alpha L^2.
\label{xiaswlz}
\end{equation}
This is obtained from the large-distance behavior of the Gaussian
correlation (\ref{gauint}): 
\begin{equation}
G_0(0,0,z\gg 1)\approx - {z\over 2\alpha L^2} .
\label{g0asy}
\end{equation}

\begin{figure}[tbp]
\includegraphics*[scale=\graphicscale]{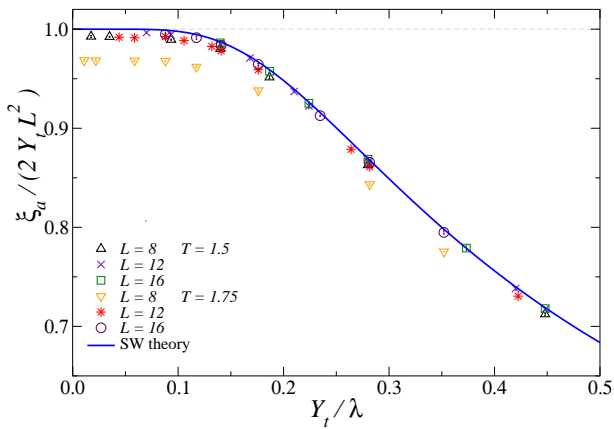}
\caption{(Color online) The ratio $\xi_a/(2Y_t L^2)$, where $\xi_a$ is
  axial second-moment correlation length defined in
  Eq.~(\ref{xiadef}), versus $Y_t/\lambda$ with $\lambda\equiv
  L_a/L^2$.  We show results from the SW theory, and QMC data for the
  3D hard-core BH model for $\mu=0$, on $L^2 \times L_a$ lattices with
  PBC (we use the QMC estimates $Y_t\approx 0.280$ and $Y_t\approx
  0.176$ respectively for $T=1.5$ and $T=1.75$).  The QMC data clearly
  approach the spin-wave AFSS with increasing $L$ (the differences get
  rapidly suppressed, apparently as $L^{-3}$).}
\label{xipfig}
\end{figure}

In order to study the AFSS of the axial coherence length, we consider
an alternative axial second-moment correlation length, as defined in
Eq.~(\ref{xiadef}).  We numerically compute it using
Eqs.~(\ref{correlation1}-\ref{gauint}), for increasing values of $L$
keeping $\lambda=L_a/L^2$ fixed, up to values of $L$ where the results
become stable with great accuracy (lattice sizes $L\gtrsim 10$ turn
out to be sufficient to get a satisfactory $O(10^{-5})$ accuracy for
the large-$L$ limit). The results show that its AFSS can be written as
\begin{equation}
\xi_a (\lambda) \approx L_a \; \tilde{f}_\xi(\zeta) = 2 \,Y_t \,L^2 
\, f_\xi(\zeta)
\label{fssxia}
\end{equation}
with $\zeta$ defined as in Eq.~(\ref{fssya}).  Again, $\tilde{f}_\xi$
and $f_\xi$ are scaling functions. In particular $\tilde{f}_\xi(0) =
0$ and $f_\xi(0) = 1$~\cite{footnotexi}.  The scaling function $f_\xi$
is shown in Fig.~\ref{xipfig}.

We also report the two-point function $G_{\rm sw}({\bf x},{\bf y})$ in
the case of open boundary conditions (OBCs), where winding effects do
not arise, but translation invariance is violated by the boundaries.
Assuming that the site coordinates are $x_\mu =
[-(L_\mu-1)/2,...,(L_\mu-1)/2]$,
\begin{eqnarray}
&&G_{\rm sw}({\bf x},{\bf y})_{\rm obc} =
\exp\left( {2M_{{\bf x}{\bf y}}-M_{{\bf x}{\bf x}} -M_{{\bf y}{\bf y}}\over 2\alpha}\right),
\qquad \label{gswobc}\\
&&M_{{\bf x}{\bf y}} = \sum_{{\bf p}\neq{\bf 0}}  2^{n_p}
\frac{\prod_{\mu} 
\cos[p_\mu (x_\mu+L_\mu/2)]
\cos[p_\mu (y_\mu+L_\mu/2)]}
{2L_1L_2L_3\,\sum_{\mu} (1 - \cos p_\mu)}
\nonumber
\end{eqnarray}
with $p_\mu = \pi \{0, \dots, L_\mu-1 \}/L_\mu$ and $n_p$ is the
number of momentum components different from zero.  Analogously to the
PBC case, for elongated $L^2\times L_a$ lattices with $L_a\to\infty$,
the phase correlation asymptotically behaves as $G_{\rm sw}(0,0,z\gg
1)_{\rm obc} \sim e^{-z/\xi_a}$ with $\xi_a$ given by
Eq.~(\ref{xiaswlz}).

Finally, assuming the universality of the above asymptotic AFSS for
any BEC system, we would like to get information on the size of the
corrections when approaching this universal limit.  Corrections to the
asymptotic scaling behavior are generally expected in generic BEC
systems due to the fact that the corresponding effective SW theories
generally require further higher-order derivative terms. Since the SW
action (\ref{swa}) is quadratic, the power law of their suppression
can be inferred by a straightforward dimensional analysis of the
couplings of the further irrelevant higher-order derivative terms
which are consistent with the global symmetries. Therefore, in the FSS
or AFSS limit, their contributions are generally suppressed by
$O(L^{-2})$, due to the fact that the next-to-leading terms has two
further derivatives, and therefore the corresponding couplings have a
square length dimension.  Therefore, we expect that BEC systems
rapidly approach the universal AFSS described by the SW quadratic
theory.  Notice that, in the presence of nontrivial boundaries,
$O(L^{-1})$ corrections may arise from boundary contributions.

\section{Space coherence in the BEC phase of the 3D BH model}
\label{becphase}

In order to check the universality of the AFSS behavior of the phase
correlations of the 3D SW theory, we consider the 3D BH model
(\ref{bhm}) in the hard-core $U\to\infty$ limit and at zero chemical
potential $\mu=0$, on elongated $L^2\times L_a$ lattices with PBC.  We
present numerical results for a few values of the temperatures below
the BEC phase transition occurring at~\cite{CN-14} $T_c\approx
2.0160$, in particular $T=1.5$ and $T=1.75$ (which are both smaller
than, but not far from, $T_c$), obtained by QMC simulations.  Some
details are reported in App.~\ref{obs}, with the definitions of the
observables considered.

At $\mu=0$ the particle density of the hard-core BH model is exactly
one half, $\rho = \langle n_{\bf x} \rangle = 1/2$, independently of
$T$ due to the particle-hole symmetry.  Moreover, the on-site density
fluctuation is exactly given by $\langle n_{\bf x}^2 \rangle-\langle
n_{\bf x}\rangle^2= 1/4$.  The connected density-density correlation
$G_n({\bf x})$, cf. Eq.~(\ref{ncorr}), turns out to differ
significantly from zero only at a distance of one lattice spacing,
where it has a negative value, while it is strongly suppressed at
larger distances, independently of the lattice size and shape.  The
corresponding values of the compressibility $\kappa = \sum_{\bf x}
G_n({\bf x})$ are $\kappa = 0.1172(1)$ and $\kappa=0.1407(1)$
respectively at $T=1.5$ and $T=1.75$.  Therefore, as expected, the
observables related to the particle density do not show any relevant
finite-size dependence within the BEC phase. As we shall see, other
observables related to the phase correlations show a more interesting
behavior.

Let us first consider the helicity modulus computed from the response
of the system to a phase-twisting field along one of the
directions~\cite{FBJ-73}, see App.~\ref{obs}.  In the case of bosonic
systems and for homogeneous cubic-like systems, it is related to the
superfluid density:~\cite{FBJ-73} $\Upsilon(T)\propto \rho_s(T)$, thus
$\Upsilon(T)$ approaches a nonzero finite value for $T\to 0$.

However, for anisotropic $L^2\times L_a$ systems we must distinguish
two helicity modulus along the transverse and axial directions:
$\Upsilon_t\equiv T Y_t$ and $\Upsilon_a\equiv T Y_a$ from twisting
along the transverse and axial directions respectively,
cf. Eqs.~(\ref{helmt}) and (\ref{winding}).  Their behaviors appear
analogous to those obtained for the 3D SW theory in elongated
geometries. Indeed, the QMC data at both $T=1.5$ and $T=1.75$ show
that $Y_t$ is stable with respect to variations of $\lambda=L_a/L^2$,
while $Y_a$ significantly decreases when increasing $\lambda$.
Straightforward large-$L$ extrapolations of the finite-$L$ data lead
to the estimates $Y_t=0.280(1)$ at $T=1.5$ and $Y_t = 0.176(1)$ at
$T=1.75$.

The QMC data for the ratio $R_Y\equiv Y_a/Y_t$ behave consistently
with the AFSS behavior (\ref{fssya}) of the 3D SW model.  They are
shown in Fig.~\ref{yafig}.  With increasing $L$, the data plotted
versus $Y_t/\lambda$ rapidly approach the SW curve for both
temperatures $T=1.5$ and $T=1.75$, supporting the universality of the
SW results.  Note that the temperature dependence enters only through
the temperature dependence of the transverse helicity modulus $Y_t$.

The convergence of the QMC data to the asymptotic universal curve is
apparently characterized by an $O(L^{-3})$ approach, which is a higher
power then the typical $O(L^{-2})$ corrections expected in SW
effective theories, see the discussion at the end of Sec.~\ref{3dsw}.
This may be explained by the fact that we are considering observables
related to the axial size that scales as $L^2$ in the AFSS limit.
Thus, in the derivative expansion of the corresponding effective SW
effective theory, a further derivative with respect to the axial space
coordinate formally leads to a further power $L^{-2}$, instead of
$L^{-1}$.

We also consider the axial second-moment correlation length $\xi_a$,
defined in Eq.~(\ref{xiadef}). Again, its large-$L$ behavior is
consistent with the AFSS obtained within the SW theory,
cf. Eq.~(\ref{fssxia}), as clearly demonstrated by the QMC data shown
in Fig.~\ref{xipfig}.

\begin{figure}[tbp]
\includegraphics*[scale=\graphicscale]{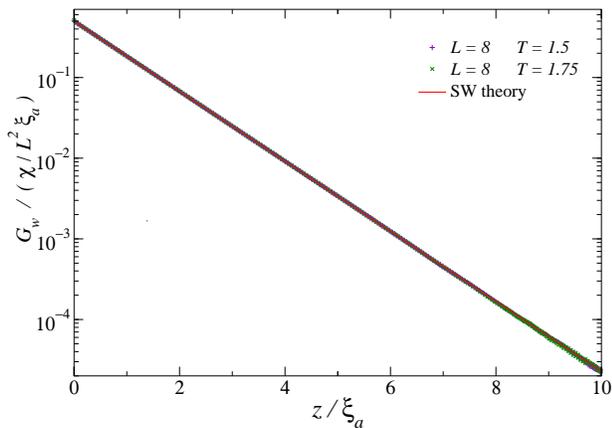}
\caption{(Color online) The wall-wall phase correlation $G_w(z)$,
  cf. Eq.~(\ref{gwcorr}), in the infinite axial-size limit
  $\lambda\to\infty$.  Again, the QMC data for the hard-core BH model
  and the computations using the 3D SW theory perfectly agree
  (actually their differences are hardly visible).  Clearly, $G_w(z)$
  decays exponentially, as $e^{-z/\xi_a}$ with $\xi_a = 2 Y_t L^2$.  }
\label{infvcopbc}
\end{figure}

Finally, we consider the space dependence of the two-point function
along the axial direction, and, in particular, of the axial wall-wall
correlation defined in Eq.~(\ref{gwcorr}).  We expect that 
its AFSS reads 
\begin{equation}
G_w(z) =  {\chi\over L^2\xi_a} \; f_w(z/\xi_a,Y_t/\lambda)
\label{fssgwz}
\end{equation}
where $\chi$ is the spatial integral of $G$.  In Fig.~\ref{infvcopbc}
we show QMC data in the infinite axial-size limit,
i.e. $\lambda\to\infty$ (obtained by increasing $L_a$ at fixed $L$, up
to the point the data become stable within errors).  Again the QMC
data nicely agree with the corresponding space dependence of the SW
two-point function, obtained using formulas
(\ref{correlation1})-(\ref{gauint}).  In particular, they confirm the
large-distance exponential decay with axial correlation length
$\xi_a=2Y_tL^2$, in agreement with Eqs.~(\ref{xiaswlz}) and
(\ref{fssxia}).

\section{Conclusions}
\label{conclu}

We have studied the equilibrium phase-coherence properties of
Bose-condensed particle systems, focussing on their shape dependence
and FSS. In particular, we consider anisotropic $L^2\times L_a$
geometries with PBC, in the AFSS limit $L\to\infty$ keeping the ratio
$\lambda\equiv L_a/L^2$ fixed.

The long-range phase-coherence properties of the BEC phase are
effectively described by a 3D SW theory, which allows us to compute
the asymptotic AFSS of the phase correlations.  Such a behavior is
universal, in particular it is independent of $T$. Indeed the
temperature dependence enters only through a normalization of the
scaling variable $\lambda$, which can be related to the helicity
modulus $\Upsilon_t$ along the transverse directions of size $L$.
Phase decoherence occurs in the limit of very elongated systems, where
the axial coherence length $\xi_a$ remains finite in the limit
$\lambda \gg 1$, and proportional to the transverse area. In
particular, in the case of PBC and for $\lambda \to \infty$, we obtain
$\xi_a=2 L^2 \Upsilon_t/T$.  Since the transverse helicity modulus
maintains its correspondence with the superfluid density,
i.e. $\Upsilon_t \propto \rho_s$, this relation may be turned into
\begin{equation}
\xi_a \propto {\rho_s\over T} A_t,
\label{xiarhos}
\end{equation}
 where $A_t$ is the transverse area.

To check the universality of the AFSS of the phase correlations, we
consider the 3D BH lattice model (\ref{bhm}), which models gases of
bosonic atoms in optical lattices~\cite{BDZ-08}.  We present a
numerical analysis in the hard-core $U\to\infty$ limit based on QMC
simulations within the BEC phase.  The results confirm the
universality of the AFSS described by the 3D SW theory. We stress that
analogous behaviors are expected for finite on-site couplings $U$, and
for generic BEC systems not constrained by a lattice structure.

The main features of the AFSS of anisotropic systems with PBC are
expected to extend to other boundary conditions (appropriate to
quantum many-body systems), in particular to the case of open boundary
conditions (OBCs) which are more realistic for experimental setup.
For example, in the infinite-axial-size limit and for OBC along all
directions, the two-point correlation function of 3D SW model,
cf. Eq.~(\ref{gswobc}), shows an exponential decay along the axial
direction as well. Analogously to the PBC case, the axial correlation
length behaves as in Eq.~(\ref{xiarhos}).

An analogous space-decoherence phenomenon has been put forward for
inhomogeneous atomic systems within elongated harmonic
traps~\cite{PSW-01}.  The bosonic systems in harmonic traps were
studied assuming the Thomas-Fermi approximation for the space
dependence of the particle density, and a Gaussian theory for the
phase fluctuations~\cite{PSW-01}.  Corresponding experimental evidence
for the BEC of harmonically trapped $^{87}$Rb atoms has been also
reported~\cite{Dettmer-etal-01,Hellweg-etal-02}.  The crossover from
3D BEC systems to the effectively 1D phase-fluctuating condensate
regime has been also discussed for ring-shaped traps at $T\ll T_c$,
using an approach based on the Gross-Pitaevskii
equation~\cite{Mathey-etal-10}, leading to analogous coherence
properties in the limit of very elongated systems.

We finally note that our results for homogeneous BEC phases should also
be of experimental relevance, since cold-atom systems constrained by
effectively homogeneous traps have been recently
realized~\cite{NGSH-15}.

It is worth mentioning that AFSS scenarios analogous to those of 3D
BEC systems are also expected in generic 3D O($N$)-symmetric
statistical systems, such as 3D $N$-vector models, within their
low-temperature phase where a spatially uniform magnetic field drives
first-order transitions~\cite{FP-85,PV-16} (in BEC systems an external
field coupled to the condensation order parameter is not physical).

\begin{acknowledgements}         
We acknowledge interesting and useful discussions with Enore Guadagnini and
Mihail Mintchev.
\end{acknowledgements}

\appendix

\section{Quantum Monte Carlo simulations of the 3D BH model and observables}
\label{obs}

In our numerical study of the 3D BH model (\ref{bhm}), we present
results obtained by QMC simulations of the BH model in the hard-core
$U\to\infty$ limit, for temperature values $T<T_c$, on anisotropic
$L^2\times L_a$ lattices with PBC, for various values of $L$ (generally
up to $L=16$) and $L_a$.  We use the directed operator-loop
algorithm,~\cite{SK-91,SS-02,DT-01} which is a particular algorithm
using the stochastic series expansion method.

The decay of phase coherence along the axial direction can be
quantified by the one-particle correlation (\ref{gbdef}).  The
corresponding length scale $\xi_a$ may be extracted from its
exponential decay or its second moment along the axial direction.  In
the case of PBC translational invariance implies
$G({\bf x - y})\equiv G({\bf x},{\bf y}) \equiv 
\langle b_{\bf x}^\dagger b_{\bf y} \rangle$.
We  define its zero-momentum component 
\begin{equation}
\chi = \sum_{\bf x} G({\bf x})
\label{chidef}
\end{equation}
and  the axial second-moment correlation length as
\begin{equation}
\xi^2_{a} \equiv  
\frac{1}{4 \sin^2(\pi / L_a)}
 \left( \frac{\widetilde{G}({\bf 0})}{\widetilde{G}({\bf p}_{a})}
                            -1  \right), 
\label{xiadef}
\end{equation}
where $\widetilde{G}({\bf p})$ is the Fourier transform of $G({\bf
  x})$ and ${\bf p}_a = (0, 0, 2\pi /L_a)$.  Moreover, we consider the
wall-wall correlation function along the axial direction, defined as
\begin{equation}
G_{w}(z) \equiv \frac{1}{L^2} \sum_{x_1, x_2} G(x_1, x_2, z).
\label{gwcorr}
\end{equation}

The helicity modulus $\Upsilon$ is a measure of the response of the
system to a phase-twisting field along one of the lattice
directions~\cite{FBJ-73}. In the case of bosonic systems and for
homogeneous cubic-like systems, it is related to the superfluid
density: $\Upsilon(T)\propto \rho_s(T)$.  However, for anisotropic
$L^2\times L_a$ systems we must distinguish two helicity modulus along
the transverse and axial directions.  In our QMC with PBC they can be
estimated from the linear winding number $W_t$ and $W_a$ along the
trasverse and axial directions respectively.  Therefore, we define
transverse and axial helicity modulus:
\begin{eqnarray}
 &&   \Upsilon_t\equiv \frac{1}{L_a} \left.
    \frac{\partial^2 F(\phi_t)}{\partial\phi_t^2}
    \right|_{\phi_t=0}  = T Y_t, \label{helmt}\\
 &&   \Upsilon_a \equiv  \frac{L_a}{L^2} \left.  \frac{\partial^2
      F(\phi_a)}{\partial\phi_a^2} \right|_{\phi_a=0} = T Y_a,
      \label{helma}\\
&&Y_t   \equiv  \frac{1}{L_a} \langle W_t^2 \rangle,
\qquad Y_a \equiv \frac{L_a}{L^2} \langle W_a^2 \rangle,
\label{winding}
\end{eqnarray}
where $F=-T\ln Z$ is the total free energy, $\phi_t$ and $\phi_a$ are
twist angles along one of the transverse directions and along the
axial direction respectively.

We also consider the particle density $\rho = \langle n_{\bf x}
\rangle$, and the compressibility
\begin{eqnarray}
&&
\kappa \equiv {\partial \rho\over \partial \mu} = {1\over V} 
\sum_{{\bf x},{\bf y}} G_n({\bf x},{\bf y}),
\label{kappadef}\\
&&
G_n({\bf x}, {\bf y}) \equiv \langle n_{\bf x} n_{\bf y} \rangle -
                      \langle n_{\bf x} \rangle \langle n_{\bf y}
                      \rangle \; .
\label{ncorr}
\end{eqnarray}


\begin{thebibliography}{99}
\bibitem{footnotexi} Although the second-moment correlation length
  generally differs from that obtained from the large-distance
  exponential decay, they are expected to be equal for Gaussian-like
  theory, as supported by the direct computations at fixed $L$ and
  $L_a\to\infty$.
\end{thebibliography}
\end{document}